# Lessons Learned in Migrating from Swing to JavaFX

Martin P. Robillard and Kaylee Kutschera

### Abstract

The authors describe a case study of the migration of an interactive diagramming tool written in Java from the Swing Graphical User Interface framework to the more recent JavaFX framework. The study distills the authors' experience identifying what information was needed to support the migration effort, and how the information was ultimately discovered. The outcome is presented in a set of five lessons about the discrepancies between expectations and reality in the search for information when migrating software between major frameworks.

## Article

Even the most risk-adverse project leaders will eventually face the question of whether to migrate to a new framework. This question can be filled with dread because the number of technical details one needs to know to master a large framework is enormous, and the number of things that can go wrong in a major migration is basically infinite.

We recently faced these challenges in the evolution of JetUML, an open-source diagram editor we develop and maintain for teaching and professional use.[1] JetUML was originally built using the Swing graphical user interface (GUI) framework. Because the amount of resources available to support the development of JetUML is minimal, one principle guiding its evolution is to minimize the risk of any event that would require heavy development effort. For this reason, the use of the more modern JavaFX GUI framework was continually put off. Ultimately, the main deciding factor for moving to JavaFX was simply the inevitability of the migration. Adapting to new hardware environments (high-resolution displays, multiple monitors) was becoming necessary, and it was inevitable that a future development would eventually render Swing-based applications obsolete.

We undertook the complete migration of the tool from one GUI toolkit to another with the perspective of both modernizing the software and learning about the major migration challenges. As is typically the case for framework migration, we had a high level of expertise in the previous framework (Swing) and only a modicum of expertise in the new framework (JavaFX). To a large extent, the migration challenge was one of information discovery. We thus construed the migration project as a case study of the questions: what information was necessary to complete the migration, and how did we discover this information? The experience led to a number of realizations about the information-seeking process involved in migrating between major frameworks.

## A Brief History of the Project

JetUML is a medium-sized, pure-Java desktop application to create and edit diagrams in the Unified Modeling Language (UML). The project started in January 2015 as an offshoot of the Violet diagram editor.[2] Although Violet was itself spun-off as an open-source project, the first author launched JetUML to focus exclusively on a minimalistic set of features intended to make diagramming as seamless as possible. The main usage scenario for JetUML is the interactive creation and modification of diagrams as part of lectures, design reviews, and other similar types of presentations. The application relies critically on its graphical user interface framework, which was originally AWT/Swing.

Before the migration, JetUML consisted of 9.1k non-comment lines of Java source code (LOCs) and 1.7k lines of comments distributed over 83 sources files organized in five packages (the data is for Release 1.2). The project was also supported by a suite of 255 JUnit tests comprising 6.3k LOCs. Figure 1 illustrates some of the salient points of JetUML's architecture related to the migration effort. From the diagram, we can distinguish three layers of architectural elements. The necessary windowing elements of the GUI framework are grouped in the layer named "Swing". These are subclassed by the application, as in most cases of framework usage, resulting in a group of elements that represent what we refer to as the windowing, or "high-level" design elements (`EditorFrame`, `GraphFrame`, and `GraphPanel`). The bottom ("low-level") layer consists of the application classes necessary to construct and draw various diagrams. Although not shown on the figure, the types `Graph`, `Node`, and `Edge` are extensively subclassed in the application with concrete elements (e.g., `ClassDiagramGraph`, `DependencyEdge`, etc.).

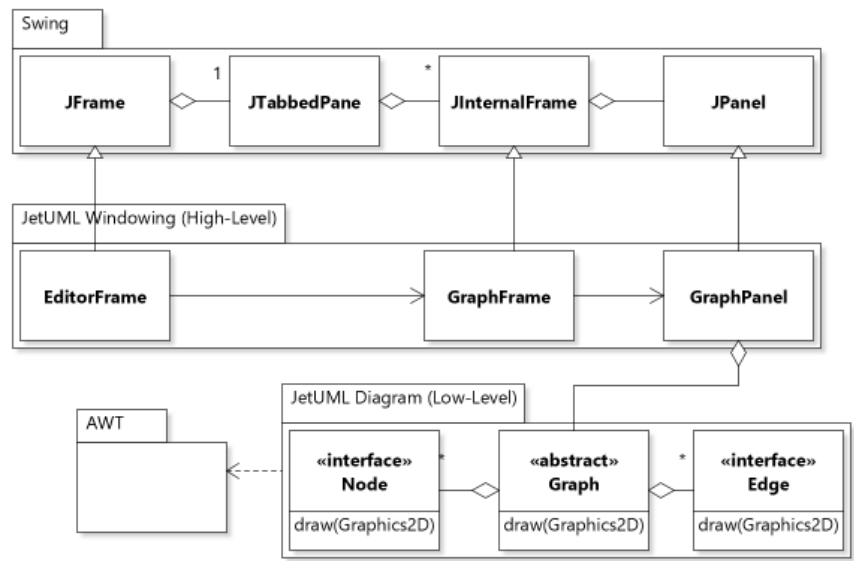

*Figure 1 Architecture of JetUML Prior to the Migration. The diagram also illustrates the output of the tool.*

## Migration Process

We organized the migration in three phases: *Preparation*, *Migration*, and *Consolidation,* with the Preparation and Consolidation phases to be completed by the main project developer (the first author) and the Migration phase to be completed by the second author. This alternation in developers between phases thus created a requirement for the design to be understandable at the end of each phase.

In the Preparation phase, the first author refactored the design to isolate as much as possible of the code that relied on the Swing framework. This refactoring involved three major efforts:

1. Separating the view from the model for diagram elements (nodes and edges);
2. Converting all references to framework-dependent geometric objects (points, lines, etc.) from Swing classes to framework-independent specific classes;
3. Replacing the framework-dependent JavaBeans-based persistence with a framework-independent solution that used the JSON notation.

Because the Preparation phase was structured as a refactoring of the existing code, we deferred research about JavaFX to the following phase. As part of the second step, we decided to convert from floating-point geometry to integer geometry in an attempt to simplify the code. The result of the Preparation phase was released as version 2.0-alpha. It brought the code base size to 12.2kLOC in 126 files with the support of 7.3kLOC of testing code in 310 tests. The 34% increase in the size of the code base was due primarily to the integration of a subset of a $3^{rd}$-party JSON processing library and the creation of numerous new classes to isolate geometry operations and separate the view from the model for diagram elements.

The main focus for the Migration phase was to migrate the code while changing as little as possible in the look and functionality of the tool, and to accomplish this goal in as incremental a way as possible given the features of the source and destination frameworks. Attempting the Migration phase is what really brought the question of information needs to a head. JavaFX is a huge framework: in the JDK 8 version there are close to 700 classes in `javafx`.* packages. With a framework this size, it is impossible to know all there is to know in advance.

Before starting the Migration phase, we searched the web for insights into the migration process and the likely problems we could run into. This investigation lead to a collection of articles, forum posts, videos, and reference documentation. Unfortunately, at that stage the technical advice proved either too specific (focusing on detailed uses cases) or too general (discussing broad issues such as threading). The real questions we faced were questions of process: how do we find out what we need to know? How do we manage the unknown? What are the traps and risk of venturing into the unknown?

Ultimately we decided to research in detail the information that would allow us to devise an overall migration strategy, and defer more detailed research until we faced concrete technical issues. The information we needed to plan our migration strategy concerned adapter components. Adapter components enable the use of components of one framework in the other framework. With both Swing-to-JavaFX and JavaFX-to-Swing adapters, it can, in principle, be possible to follow either a strategy of top-down migration (migrated JavaFX windows containing legacy Swing widgets) or bottom-up migration (legacy Swing windows containing migrated JavaFX widgets), or any combination of the two that isolates changes and limits risk.[6] In the end we selected a top-down migration strategy so that we could migrate the more stable ("architectural") part of the design first, and defer the migration of the drawing code, which required more uncertainty and experimentation, to later, when the JavaFX-supported window structure was in place. The result of the migration phase was released as version 2.0, which was almost identical to 2.0a in terms of size metrics.

Finally, the idea of the Consolidation phase was to solidify the migrated version with various cosmetic improvements, design simplifications, and adaptations made directly possible by JavaFX. The result of this phase was released as version 2.1, with 12.3kLOC in 142 files, with additional improvements committed to version 2.2 and 2.3. The complete code base of all releases of JetUML can be obtained from its GitHub repository.[1] In the end we successfully completed the migration at the cost of approximately 3 person-months, with additional rework after the initial JavaFX release.

## Lessons Learned

The migration process led to five major realizations and corresponding implications about looking for technical information to support a framework migration. Although the realizations are derived from our experience migrating to JavaFX, they are not tied specifically to details of JavaFX, so we can expect that they will generalize to situations where frameworks exhibit similar characteristics. For each lesson, we present a summary of the technical details and a discussion of the information-seeking context, realization, and how we would have leveraged each lesson, had we known it in advance.

### Adapting to Detail

Our first lesson concerns the information we expected to be easy to find but for which the easily found resources were largely insufficient. The cornerstone of our migration plan was to use adapter components to support an incremental process. Adapters are a classic strategy for migrating from one framework to another[4, 5] so we knew what to look for. Adapter components are available in both Swing and JavaFX[10]. Swing includes a `JFXPanel` class to embed JavaFX components into a Swing application, and JavaFX includes a `SwingNode` class to embed Swing components into a JavaFX application. Both classes provide corresponding API documentation. Additionally, the official Java documentation site includes an article on JavaFX-Swing interoperability.[10] With this information in hand, it appeared the use of components was a straightforward task. One popular answer on an on-line forum even claims that adapters make migrations to JavaFX easy.[7]

Unfortunately, when we attempted to use the adapters, numerous practical issues surfaced that required additional insights and showed the glaring limitations of the official documentation, which only showed how to use the adapters in basic usage scenarios, and provided insights on only one additional concern, concurrency.

*Performance*: Top-down migration requires the use of the `SwingNode` adapter class, which can hold Swing content in JavaFX windows. This strategy leads to performance issues because `SwingNode` instances are not meant to hold heavyweight components. In contrast, bottom-up migration using the `JFXPanel`, does not have these problems. Hybrid migration strategies, such as interposing a JavaFX component between Swing components, can result in major performance problems such as large delays when first loading a window.

*Computing Dimensions*: Because components in one framework are embedded in the other, we found that it was not possible to properly compute the preferred sizes of components from both frameworks. To address sizing problems, one recommendation was to hard code fixed preferred sizes until they can be properly computed by the framework[6], which adds development overhead.

*Dependency Cycles*: During migration, it may be necessary to have cyclic dependencies between classes if it is necessary to access a parent component. These dependencies can be removed once child and parent are contained in the same framework and can access each other through the scene graph. For example, in JetUML, because we did a top-down migration, a high-level windowing element (the tabbed pane) was migrated to JavaFX before the drawing area it contains, which remained a Swing component. A reference to a diagram's tab was needed by the Swing drawing area to update the tab's title. The JavaFX parent component was not accessible in the Swing child because there is no way to access the `SwingNode` instance that the child Swing component is wrapped in.

In the end, the **problem** we faced in this case is that information was easy to find but inevitably shallow because the use of adapters is very context-sensitive. The **consequence** is that we had to spend much effort hunting for scarce experience reports[6] and experimenting to determine when to use the components, despite having easily found highly relevant documentation. The **lesson** is that existing documentation on how to use a pivotal component in a small, synthetic scenario is bound to be incomplete. In hindsight, it would have been better to make this realization earlier and invest time in prototyping and experimentation with the components in context. Ideally, though, documentation for adapters would benefit from a list of the main implications for using them.

**False Friends**

In linguistics, the expression "false friend" refers to a term in a language that is deceptively similar to a term in a different language, while having a different meaning. For example, the term "billion" commonly refers to $10^9$ in English but $10^{12}$ in French and German. As we realized, the same phenomenon occurs in the context of translation between frameworks.

A major concern for a diagram editor is to draw shapes. Swing supports this functionality through a `Shape` class hierarchy, with subclasses such as `Arc2D`, `Ellipse2D`, etc. In Swing, a `Shape` instance can be drawn on a graphics context simply by calling `context.draw(Shape)`. In our preliminary investigation of JavaFX, we had noticed that it defined a near-equivalent API, with also a class `Shape` with equivalent subclasses with the same name (except the `2D` suffix). This realization initially gave us the impression that migrating the drawing code would be a trivial exercise in mechanical translation, and the exact correspondence of names even made the need for advanced API migration mining tools superfluous.[3] Unfortunately, the feeling of elation was shattered when we realized that in JavaFX the graphics context object does not have a method to draw `Shape` instances, and that in fact in JavaFX, `Shape` instances are not used to draw shapes directly, but rather for a new purpose that did not exist in Swing (to place shapes in a scene). Consequently, the code had to be extensively refactored to adapt our former strategy (to create a `Shape` instance in various diagram element classes and draw it once), to one that was supported by JavaFX (namely, to draw shapes in each diagram element class using available drawing primitives).

The **problem** in this case is that by noticing an API type hierarchy with near-identical names, we made the false assumption that we did not need to invest additional effort in information search for the corresponding part of the migration. The **consequence** is that we needed to change the original design after the migration was partially completed. The **lesson** we learned in this case is simple: beware of false friends. A basic procedure that would have avoided the deception would have been to inspect references to (Swing) `Shape` classes to ensure the same services were available in JavaFX. As it turns out, they were not.

**Feature Gap**

A large influence on the design of the Swing version of JetUML was what seemed originally like a modest feature: the ability to draw shapes directly on user interface components. In Swing, most user interface components are subclasses of `JComponent`, which defines a `paintComponent(Graphics)` method. Subclasses can then override this method to draw shapes directly on the component. We leveraged this simple feature to orchestrate the visual rendition of diagrams. Figure 2 summarizes the key aspects of this design. The main point of the figure is the simplicity of the design: when it determines that a refresh operation is necessary, the framework simply calls `paintComponent`, which delegates the drawing of the graph. The code in `graph.draw(…)` then calls drawing primitives directly on the `Graphics` object.

The biggest hurdle we faced in our migration project is when we realized that it was not possible to draw directly on UI components in JavaFX, and that a major redesign of the drawing mechanism was thus necessary. Unfortunately, we found no information on how to handle this migration situation. The ability to draw shapes obviously exists in JavaFX. However, because the mechanism was redesigned, there is no straightforward migration path to be followed (e.g., replacing a class with another). We conjecture that because the number of ways to adapt to the feature gap is open-ended and context-sensitive, it was not feasible to document specific migration paths. In any case the situation created two major sources of information needs for moving forward: 1) discovering viable migration paths and selecting one; 2) discovering all the technical information required to implement the selected path.

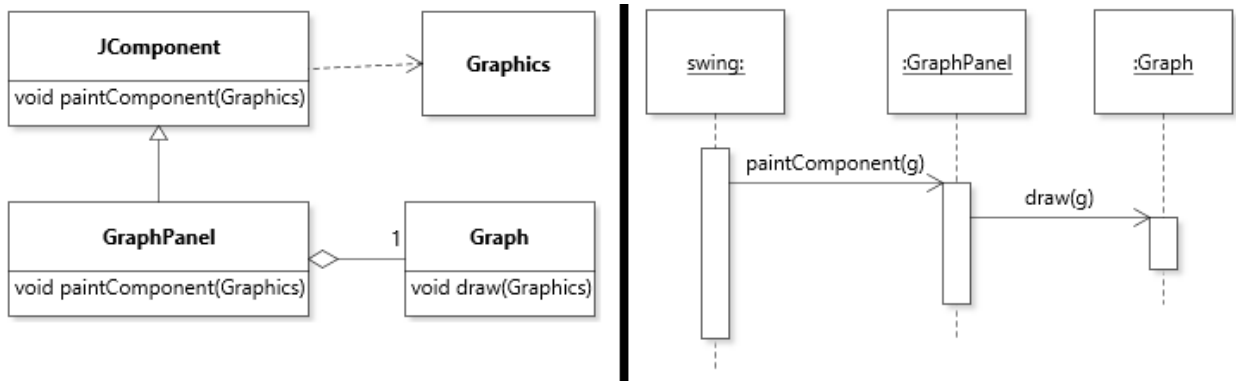

*Figure 2 Design of the diagram drawing concern.*

In JavaFX, the creation of drawings can be done in two different ways that operate at two different levels of abstraction. The high-level mechanism involves creating instances of class `Shape` and adding them to a `Pane`, so as to constitute a collection of drawable objects. The low-level mechanism involves creating a drawing directly on a `Canvas` instance by using drawing primitives on the canvas's graphics context, and embedding the canvas in a pane. Each of these mechanisms forms a somewhat polarized version of Swing's original drawing approach, which combined elements of both. After extensive experimentation to fulfill the first information need, we determined that using a scene to represent a drawing was too inefficient, and we opted for the embedded canvas solution. Unfortunately, trying to implement the canvas-based migration of the drawing feature turned out to be frustrating because technical issues caused by the difference in drawing mechanisms kept turning up and had to be researched extensively.

As an illustration, the worst issue was that in the Swing-based version of JetUML, the drawing area is resizable. To match the original feature, we tried to make the canvas resizable. Technically, this only requires overriding a few methods. Despite a number of resources documenting straightforward solutions[12], none were applicable. In experimenting with a resizable canvas, we ran into numerous layouting and sizing issues when trying to integrate it with a scroll pane  (a component that allows scrolling areas larger than the window size). Ultimately, one of the nicest features of JetUML, a diagram space that seamlessly adapts to the window size, could no longer be reasonably supported: we gave up trying and ended up investing a considerable amount of redesign effort in rethinking how the tool would work with a fixed diagram size.

It can be expected that not all features of a framework will be ported, at least initially. This is not an issue when there is a straightforward migration path or workaround. For example, JavaFX does not provide any functionality to save an image to a file. However, there is a well-documented workaround, namely to convert a JavaFX image to a Swing image using a provided utility method and then using the legacy API call to save the image. In the case of the drawing feature, the **problem** was not only the lack of a formerly-supported feature in JavaFX, but also the absence of a canonical migration path for re-implementing the feature. The **consequence** was not only a requirement to re-design the feature, but also a stark lack of technical information on how to re-implement the former functionality. In hindsight, the main **lesson** is to be aware of the critical features of the original framework we rely on. Because in our case this feature was so seamlessly supported by the framework, it had not been conspicuous as a functionality to experiment with during the Preparation phase. In the future we would try to identify important features in the abstract, independently of their implementation and support by the framework.

**Feature Blindness**

Another issue we faced related to our search for information was to explicitly avoid looking for information when we should have. One of our explicit risk-management strategy in the Migration phase was to change as little as possible in the look and functionality of the application. For this reason, our search for information focused on discovering migration paths that were literally at the code statement level. However, through later code inspection, we realized that this strategy had been overly aggressive, because we had re-implemented various features that had been custom-built in the Swing version simply because no other option had been available.

As one example, one of the custom GUI components we had built for the Swing version was a tool bar. A tool bar is a container for a button group that allows activating or selecting "tools" (such as to create a node in a diagram). In modern GUI applications, tool bar components can reorganize or hide some tools in the toolbar when the default layout does not fit in the display. Swing did not offer this feature, which is why it was custom-built. As part of our migration process, we re-built the custom version in JavaFX, only to realize all the desired functionality was available from JavaFX's `Toolbar` component.

Another example is the implementation of a "drop shadow" effect (visible in Figures 1 and 2). Swing did not have explicit support for adding drop shadows to image elements, and a lot of code was necessary to compute additional shapes and bounding boxes that included a drop shadow for all different types of diagram elements. However, with JavaFX, drop shadows can be added with a few lines of code.

In sum, the **problem** we faced is that by insisting on a literal migration, we overlooked many opportunities for simplifying the code. The **consequence** is twofold: first, we spent a lot of effort migrating complex code that could be simplified, and second, we eventually spent additional effort implementing the improvement as part of later releases. To be sure, identifying features that we can leverage as part of the migration is not free. One problem is that although knowledge of what we need as posteriori is obvious, before the migration it is much harder to know what new features can reasonably be used in the migrated version. Ultimately, the **lesson** we drew is not to be too strict about the migration. Although this is easy to say after the fact, we hypothesize that with a minimum of a priori requirements analysis, it would have been possible to identify the two most obvious features, namely the tool bar component and drop shadow effect.

**Hidden Information**

After migrating the drawing feature we noticed that the formerly razor-sharp rendering of diagrams we had experienced with Swing had been replaced by somewhat blurry diagrams. We initially blamed the problem on over-aggressive aliasing and set the issue aside for the initial JavaFX-based release 2.0. After the release and extensive experimentation on different displays, we concluded that the blurriness of diagrams was a major step back, and further investigated the issue. This investigation revealed that in the JavaFX framework "At the device pixel level, integer coordinates map onto the corners and cracks between the pixels and the centers of the pixels appear at the midpoints between integer pixel locations."[8] This means that to have a point exactly map to a pixel and render sharply, this point needs to have coordinates (0.5, 0.5). This crucial piece of information is unfortunately buried in one paragraph of the class-level documentation for class `Node`. Even more confusing, a different paragraph in the class-level documentation for `Shape` describes the blurriness problem exactly, but the solution described is inapplicable for applications that use the low-level (canvas-based) drawing mechanism. In the latter case, it is the insight in class `Node` that applies, even though in that case no mention is made of blurriness. A Stack Overflow post[11] turned out to have been instrumental in helping us assemble the solution to this puzzle from disparate pieces.

In this case, the **problem** is that information about an important technical concern with pervasive implications was not located where we would have expected in (namely an overview page, tutorial, or migration guide). Rather, it was placed in API documentation, a location we normally access as a reference when trying to answer detailed technical questions. The **consequence** is that instead of integrating this key piece of information into the migration process, it had to be used post hoc in recovery efforts. Specifically, we solved the problem in a later release by directing all shape drawing requests through methods that simply shifted the original integer coordinates by 0.5 in each dimension. The **lesson** is to deeply investigate at least the specific implications of design decisions that are related to the migration. Although we were unaware of the floating-point pixel geometry issue, we had decided to move to integer geometry. In hindsight, it would have been a good idea to further investigate the ramifications of this design decision.

## Conclusions

Most of the challenges we faced were challenges of information discovery. To migrate the code from Swing to JavaFX while maintaining high code and design quality standards required a number of key insights which we did not have in advance. Most of these insights could have been obtained through additional investigation and experimentation. However, before undertaking the migration, the task of discovering all relevant information seemed unsurmountable: there is was too much information to peruse, and our information needs were too vague.

Fulfilling the information needs of software developers is an active research area in software engineering. The article "Patterns of Knowledge in API Reference Documentation" provides a structured review of the foundational work in this area.[14] In certain cases, technology may help surface useful information, for example by discovering insightful sentences in forums like Stack Overflow,[9] or repackaging technical information into a more convenient format.[13] However, solutions that rely on existing documentation are dependent on the quality of this documentation, and do not address situations where information needs must be fulfilled through experimentation.

The lessons described in this article enable a more general solution: a structured approach for identifying information needs related to a migration. Each of the lessons provides a general context where migration information was challenging to obtain, examples of the potential consequences of insufficient research in that context, and insights on how to better structure the information search in such a context.

## Acknowledgements

The authors are grateful to Sebastian Baltes, Mathieu Nassif, Christoph Treude, and the anonymous reviewers for insightful feedback.